\documentclass[sigconf, nonacm]{acmart}
\usepackage{booktabs}
\usepackage{multirow}
\usepackage{enumitem}
\usepackage{listings}
\usepackage{cleveref}
\newcommand\vldbyear{2026}
\newcommand\vldbworkshop{}
\newcommand\vldbauthors{\authors}
\newcommand\vldbtitle{\shorttitle} 
\newcommand\vldbavailabilityurl{URL_TO_YOUR_ARTIFACTS}
\newcommand\vldbpagestyle{plain} 
\usepackage{subcaption}

\usepackage{booktabs}
\usepackage{multirow}

\usepackage{amsmath}
\usepackage{listings}
\usepackage{xcolor}

\usepackage{xcolor}
\usepackage{listings}

\begin{document}
\title{iFVS: Towards Instance-Optimized Filtered Vector Search}

%%
%% The "author" command and its associated commands are used to define the authors and their affiliations.
\author{Abdullah Al-Mamun}
\affiliation{%
  \institution{Purdue University}
  \city{West Lafayette}
  \country{USA}}
\email{mamuna@purdue.edu}

\author{Yeasir Rayhan}
\affiliation{%
  \institution{Purdue University}
  \city{West Lafayette}
  \country{USA}}
\email{yrayhan@purdue.edu}

\author{Walid G. Aref}
\affiliation{%
  \institution{Purdue University}
  \city{West Lafayette}
  \country{USA}}
\email{aref@purdue.edu}

\begin{abstract}
Filtered vector search (FVS) is increasingly important in modern AI + DB systems, where vector similarity search is combined with relational predicates. 
Quantization 
% also 
plays a vital role in these systems by enabling 
% processing queries 
query processing 
over large vector datasets. However, lossy approaches, e.g., Product Quantization (PQ), 
% also 
incur a precision penalty during distance calculation, thereby negatively impacting the query recall performance. This problem becomes more challenging in FVS because the relevant vector space can change with the relational predicate and selectivity. 
Motivated by the success of instance-optimized database system components, 
% these observations
% we study FVS through the lens of instance optimization. 
we introduce {\em iFVS}, an Instance-Optimized Filtered Vector Search technique. 
Given a fixed, quantized vector dataset, and a representative workload of filtered vector queries, {\em iFVS} adopts a query-specific codebook generation approach for FVS that is instance-optimized towards a certain dataset and query workload.
%over quantized data.
Instead of using a fixed codebook for all queries, 
{\em iFVS} conditions distance estimation on both the query vector and the filter predicate. This enables more accurate ranking over compressed vectors while preserving compact per-vector storage.
%AAM: the last sentence is a placeholder for now
Experiments show that 
%our approach 
{\em iFVS}
improves the Queries Per Second (QPS)-recall tradeoff across several filter selectivity bins compared with fixed-codebook quantized FVS baselines.

\end{abstract}

\maketitle

\pagestyle{\vldbpagestyle}
\begingroup\small\noindent\raggedright\textbf{VLDB Workshop Reference Format:}\\
\vldbauthors. \vldbtitle. VLDB \vldbyear\ Workshop: \vldbworkshop.\\ %\vldbvolume(\vldbissue): \vldbpages, \vldbyear.\\
%\href{https://doi.org/\vldbdoi}{doi:\vldbdoi}
\endgroup
\begingroup
\renewcommand\thefootnote{}\footnote{\noindent
This work is licensed under the Creative Commons BY-NC-ND 4.0 International License. Visit \url{https://creativecommons.org/licenses/by-nc-nd/4.0/} to view a copy of this license. Copyright is held by the owner/author(s). \\
\raggedright Proceedings of the VLDB Endowment. %, Vol. \vldbvolume, No. \vldbissue\ %
ISSN . \\
%\href{https://doi.org/\vldbdoi}{doi:\vldbdoi} \\
}\addtocounter{footnote}{-1}\endgroup

% \ifdefempty{\vldbavailabilityurl}{}{
% \vspace{.3cm}
% \begingroup\small\noindent\raggedright\textbf{VLDB Workshop Artifact Availability:}\\
% The source code, data, and/or other artifacts have been made available at \url{\vldbavailabilityurl}.
% \endgroup

\section{Introduction}
Modern database systems store and query high-dimensional vector embeddings for similarity search operation. Both specialized 
vector databases (e.g., Milvus~\cite{wang2021milvus}) and integrated vector databases (e.g., pgvector~\cite{pgvector, liu2026fast}) 
%vector databases 
play a vital role in the Large Language Models (LLMs) inference step~\cite{pan2024survey}. These vector embeddings 
% might 
represent a diverse set of objects (e.g., documents, images).
%AAM: introducing PQ here
Moreover, compressed representations of vector embeddings are often required for very large datasets. A common approach is to leverage quantization~\cite{gray2002quantization} techniques, e.g., Product Quantization (PQ)~\cite{jegou2010product}. PQ can reduce memory requirements by splitting vectors into separate subspaces and quantizing each subspace independently. For example, a 128-dimensional vector can be split into 16 subspaces of 8 dimensions each, where each subspace has its own independent set of learned centroids in the form of a codebook.

%Although the most common query workload is Approximate K-Nearest Neighbor search (ANN, for short),
%search, 
In many real-world scenarios, Approximate K-Nearest Neighbor search (ANN, for short) is combined with relational predicates (e.g., $=, \geq, \leq$ ). For example, a researcher might want to filter with dates while searching similar research papers/documents. 
This setup is commonly known as Filtered Vector Search (FVS)~\cite{chronis2025filtered, liang2024unify, patel2024acorn,ait2025rwalks, sievefvslizhao25,zuo2024serf,mohoney2023high,gollapudi2023filtered}. The general form of an FVS query~\cite{zhang2026vecbench} is shown below, where filter predicates are applied to col1 and col2, and ANN search is performed over vector\_col using the query vector q1.
% YR: Once you fix the abstract, this should settle nicely with no spacings
\captionsetup[lstlisting]{skip=-1pt}
\begin{lstlisting}[language=sql,basicstyle=\ttfamily\small,breaklines=true,numbers=none,tabsize=2,keywordstyle=\color{blue},aboveskip=2.5pt,belowskip=0pt,xleftmargin=0pt,framexleftmargin=0pt, caption={Example of an FVS query.}, captionpos=b,]
SELECT * FROM T1
WHERE col1 > value1 AND col2 = value2
ORDER BY distance(q1, vector_col)
LIMIT K
\end{lstlisting}

FVS imposes additional challenges to 
%the 
ANN query processing because the filter predicate changes the effective search space~\cite{chronis2025filtered}. Particularly, queries with highly-selective predicates %(i.e., low-selectivity) 
%predicate 
return a small fraction of vector data as candidates for similarity search and 
%forces employ 
uses 
brute force vector search over the candidates instead of building 
%an index 
a vector index 
per query~\cite{song2026favor}. Existing systems 
%therefore 
use multiple FVS strategies, 
%such as 
e.g.,
pre-filtering, post-filtering, in-filtering, and expanded filtering as no single strategy dominates across all filter selectivities~\cite{zhang2026vecbench}. As a result, practical systems often rely on dynamic FVS strategy selection during query processing.

Recent progress in instance-optimized systems, 
e.g.,~\cite{ding2022sagedb, stoian2150instance, ding2021instance},
has demonstrated that database systems can benefit from tailoring internal components to given data and query workloads, e.g., learned indexes~\cite{kraska2018case, al2025survey}. 
Inspired by this direction, 
in this paper, we study FVS through the lens of instance-optimization and 
%ask 
investigate 
whether the quantization component in FVS can also be instance-optimized. In particular, rather than relying on a single filtered-query-agnostic PQ codebook, we study whether the codebook can be adapted to a fixed vector dataset and a representative 
% filtered-
FVS-query workload with the goal of improving the QPS-recall tradeoff over its workload-agnostic counterparts. Although there are a few studies on query-aware quantization~\cite{jaiswal2022ood,zhang2023query}, these methods have not been studied in the context of FVS. 

In this paper, we propose a filtered-query-aware PQ codebook generation framework for FVS. We refer to this approach by {\em iFVS}, short for Instance-optimized Filtered Vector Search. Given a query vector and a relational predicate,
{\em iFVS} generates a filter-aware PQ codebook by perturbing a fixed base codebook.  The perturbation is produced from the query-predicate pair using a shared memory bank and learned adjustment directions. Moreover, the relational predicate is encoded into a filter-aware weight vector that re-weights vector dimensions according to their relevance to the filter predicate. During query processing, the filter predicate first determines the eligible candidate pool. The candidates are ranked using the generated filter-aware codebook and weights. Thus, 
%our approach 
{\em iFVS}
keeps the compact PQ codes and the base codebook fixed, while adapting the scoring representation to each filtered query.
%}

Notice that PQ is lossy and therefore approximates the true distance from the original vector using either Asymmetric Distance Computation (ADC) or Symmetric Distance Computation (SDC)~\cite{jegou2010product}. As a result, lossy quantization methods,
e.g.,
%like 
PQ,
can suffer from lower query recall compared with their full-precision counterparts. The standard PQ codebook is both query- and filter-agnostic in the context of FVS. This creates an opportunity for instance optimization, which is what iFVS does. Rather than using a single fixed codebook for all queries, iFVS adapts the original codebook to a fixed dataset and a representative filtered-query workload. 
%This paper introduces this approach that we term: Instance-Optimized Filtered Vector Search, or {\em iFVS}.

% YR: I think 3 is fine
%\abdullah{
In summary, this paper makes the following contributions:
%--}
\begin{enumerate}[leftmargin=*,label=\textbf{C\arabic*.}]
\item We initiate the study of FVS through the lens of instance-optimization.
\item We propose {\em iFVS}, a query-specific, filter-aware PQ codebook generation framework for FVS.
\item Extensive experimental evaluation on SIFT1M and SIFT10M datasets demonstrates the potential of 
%our proposed method. 
{\em iFVS.}
\end{enumerate}

%\section{Our Approach} 
%\section{iFVS}
\section{\texorpdfstring{\lowercase{i}FVS}{iFVS}}
Given a query vector $q \in \mathbb{R}^{D}$ and a relational predicate $f$, 
%we answer 
{\em iFVS} processes 
an FVS query as follows. The query-predicate pair, $\langle q, f\rangle$ is fingerprinted via $h$ independent hash functions into a shared memory bank $\mathcal{M}$ of $B$ rows. The resulting $h$ coefficient vectors that are summed to obtain a query-specific adjustment $\alpha \in \mathbb{R}^{M \times r}$.  $M$ denotes the number of sub-spaces, which is tied to the underlying PQ Codebook. $r$ denotes a hyper-parameter specifying the number of directions the adjustment can take place. The relational predicate $f$ is encoded by a light-weight Transformer into a filter-aware weight vector $w_z \in \mathbb{R}^{D}$. $w_z$ reweighs the dimensions of the query and the database vectors according to their relevance to Predicate $f$. 
% \begin{equation}
% \alpha = w(z)
%     + \sum_{i=1}^{k} \texttt{bank}\!\left[\texttt{hash}_i(\text{id})\right]
% \end{equation}
The adjustment $\alpha$ is combined with a learned parameter $W \in \mathbb{R}^{M \times r \times K \times d}$ to produce 
%to produce 
the query-specific perturbation of the PQ Codebook, making it filter-aware. $K$ denotes the number of codewords per subspace, and $d$ denotes the dimensionality of each codeword in the PQ codebook. Together, $\alpha$ and $\mathbf{w}_f$ define 
%our 
{\em iFVS}
perturbed filter-aware Codebook $C_{\text{iFVS}}$, which is built on top of the PQ Codebook $ C_{\text{PQ}}$. 
\begin{align*}
\boldsymbol{\Delta} &= \alpha \cdot W \qquad w_f = \text{softplus}\!\left(w_z\right)\\
C_{\text{iFVS}} &= C_{\text{PQ}} + \boldsymbol{\Delta}
\end{align*}

Given $C_{\text{iFVS}}$, our approach executes an FVS-query in the following three stages.
\begin{enumerate}[leftmargin=*,label=\textbf{\arabic*.}]
    \item \textbf{Filter.} The relational predicate $f$ is evaluated over each vector node's relational attributes to identify a candidate pool $\mathcal{P}$.
    \item \textbf{Rank.} For each candidate vector$v\in\mathcal{P}$, we reconstruct its approximate vector from our filter-aware codebook $C_{\text{iFVS}}$. The candidate is then scored as follows. 
    \begin{align*}
      \text{score}(v) &=
      \sum_{m=1}^{M} \sum_{d=1}^{D/M}
      \Bigl[
        2\,(\mathbf{w}_f)_{m,d}\, \mathbf{q}_{m,d}\, C_{\text{iFVS}}[m, \text{code}_i[m]]_d\\
        &\;-\;
        (\mathbf{w}_f)_{m,d}\, C_{\text{iFVS}}[m, \text{code}_i[m]]_d^{\,2}
      \Bigr]
  \end{align*}
    \item \textbf{Select.} We return the top-$k$ candidates ranked by score.
\end{enumerate}

\noindent \textbf{Learning $C_{\text{iFVS}}$ and $w_f$.} 
At the start of training, we initialize $\alpha = 0$ and $w_f = 1$. Thus, the model initially reduces to the original PQ codebook, i.e., $C_{\text{eff}} = C_{\text{PQ}}$, with no predicate-aware reweighting. From this initialization, we train $\alpha$ and $w_f$ to make the PQ representation filter-aware. For each training query-predicate pair $\langle q,f\rangle$, the model constructs the current $C_{\text{iFVS}}$ and $w_f$, as described above. It then scores the ground-truth positive vectors and mines hard negatives from the candidate pool $\mathcal{P}$. Precisely, hard negatives are the highest-scoring non-ground-truth candidates under the current $C_{\text{iFVS}}$ and $w_f$, since these are the candidates most likely to be confused with the true answers. We also add an anchor regularizer that keeps  $C_{\text{iFVS}}$ close to the original $C_{\text{PQ}}$. During training, gradients flow through $\alpha$ and updates the memory bank $\mathcal{M}$, the learned parameter $W$, and $w_f$. The base codebook $C_{\text{PQ}}$ and the PQ codes remain fixed throughout training.

\begin{table*}[t]
\centering
\small
\setlength{\tabcolsep}{4pt}
\renewcommand{\arraystretch}{0.85}
\caption{%
  Index construction and search parameters for SIFT1M and SIFT10M datasets.
  $M$: HNSW graph connections per layer;
  $\mathrm{ef}_{\mathrm{build}}$: beam width during construction;
  $n_{\mathrm{list}}$: number of IVF inverted lists.
  Search parameters $\mathrm{ef}$ (HNSW) and $\mathrm{nprobe}$ (IVF) are swept over
  the listed ranges to trace QPS-recall tradeoff curves;
  IVFFlat and IVFPQ share identical $\mathrm{nprobe}$ ranges.
  The same parameter ranges are used for both post-filtering and in-filtering.
  ``---'' denotes not applicable.%
}
\label{tab:index_params}
% \resizebox{\textwidth}{!}{%
\begin{tabular}{@{}ll ccc ccc@{}}
\toprule
\multirow{2}{*}{Dataset}
  & \multirow{2}{*}{Index}
  & \multicolumn{3}{c}{Construction Parameters}
  & \multicolumn{3}{c}{Search Parameters} \\
\cmidrule(lr){3-5}\cmidrule(lr){6-8}
  & & $M$ & $\mathrm{ef}_{\mathrm{build}}$ & $n_{\mathrm{list}}$
  & $s = 0.01$ & $s = 0.05$ & $s = 0.10$ \\
\midrule
\multirow{3}{*}{SIFT1M}
  & HNSWFlat
    & 16 & 200 & ---
    & ef: $100$--$5000$
    & ef: $100$--$2000$
    & ef: $100$--$1000$ \\
  & HNSWPQ
    & 16 & 200 & ---
    & ef: $100$--$20000$
    & ef: $100$--$10000$
    & ef: $100$--$5000$ \\
  & IVFFlat / IVFPQ
    & --- & --- & $4096$
    & nprobe: $16$--$1024$
    & nprobe: $16$--$1024$
    & nprobe: $16$--$1024$ \\
\midrule
\multirow{3}{*}{SIFT10M}
  & HNSWFlat
    & 32 & 400 & ---
    & ef: $100$--$10000$
    & ef: $100$--$5000$
    & ef: $100$--$2000$ \\
  & HNSWPQ
    & 32 & 400 & ---
    & ef: $100$--$20000$
    & ef: $100$--$10000$
    & ef: $100$--$5000$ \\
  & IVFFlat / IVFPQ
    & --- & --- & $16384$
    & nprobe: $64$--$1024$
    & nprobe: $64$--$1024$
    & nprobe: $64$--$1024$ \\
\bottomrule
\end{tabular}%
% }
\end{table*}

\section{Evaluation}

We run all experiments on an Ubuntu 18.04 with Intel Xeon Platinum 8168 (2.70GHz) and 3TB of total available memory. Moreover, the search performance is evaluated under single-threaded execution using the FAISS~\cite{douze2025faiss} library. Both training and inference for iFVS run on an NVIDIA
A30 Tensor Core GPU.

\subsection{Dataset and Filter Query Workloads}
We use the SIFT1M and SIFT10M datasets~\cite{texmex} for all experiments. Each dataset contains 128-dimensional base vectors and 10K query vectors without any filtering conditions. For the filtered query workloads, we generate filtered vector-search queries by imposing relational predicates over a fixed set of SIFT dimensions. The filterable attributes are selected to cover different spatial cells of the SIFT descriptor. For each query and each selectivity bin, the generator samples one predicate over one to three of these fixed dimensions. We use three target selectivity bins for the filter predicates: [0.01, 0.05, 0.10]. For each selectivity bin, we use the 10K SIFT query vectors as the base query set and assign generated filter predicates to each query. This creates a filtered workload in which every ANN query is evaluated only over the subset of base vectors that satisfy its predicate. Finally, we compute the exact top-100 filtered ground truth directly over the predicate-passing vectors.

\subsection{Baselines}
We use 13 baselines for our evaluation where 12 baselines are standard FAISS implementation except the Pre\_SDC\_numpy which has been implemented using numpy. For the post- and In-filtering baselines, we have included the HNSW~\cite{malkov2018efficient} and IVF~\cite{jegou2010product} indexes. For theh PQ-encoded dataset, we choose 16 subspaces of 8 dimensions each. Table~\ref{tab:index_params} presents the construction and search parameters of the baseline indexes for SIFT1M and SIFT10M.

%AAM: Compacted to save space
\noindent\textbf{B1. Pre-filtering.} These methods evaluate the filter predicates first to obtain the candidate set, then perform the distance computation over that set of vectors:
%\begin{itemize}[leftmargin=*]
%\item \textbf{\texttt{Pre\_IndexFlatL2}} 
(i) Pre\_IndexFlatL2 
performs exact L2 distance calculation over the full-precision vectors.
%\item \textbf{\texttt{Pre\_IndexFlatL2\_PQ}} 
(ii) Pre\_IndexFlatL2\_PQ 
reconstructs vectors from their PQ codes and performs exact L2 distance calucation.
%\item \textbf{\texttt{Pre\_IndexPQ\_ADC}} 
(iii) Pre\_IndexPQ\_ADC 
ranks the candidate set over quantized data with Asymmetric Distance Computation (ADC).
%\item \textbf{\texttt{Pre\_IndexPQ\_SDC}} 
(iv) Pre\_IndexPQ\_SDC 
ranks the candidate set over quantized query + data with Symmetric Distance Computation (SDC).
%\item \textbf{\texttt{Pre\_SDC\_numpy}} 
(v) Pre\_SDC\_numpy 
is a NumPy-based implementation of pre-filtered SDC scoring over PQ-encoded vectors.
%\end{itemize}

\noindent\textbf{B2. Post-filtering.} These indexes run the vector search first and then discard results that do not satisfy the filter predicates: \texttt{Post\_HNSWPQ}, \texttt{Post\_IVFPQ}, \texttt{Post\_HNSWFlat}, and \texttt{Post\_IVFFlat}.

% \vspace{2pt}
\noindent\textbf{B3. In-filtering.} These indexes integrate the filter predicate directly into the search by providing a filtering bitmap, skipping non-qualified candidates during traversal: \texttt{In\_HNSWPQ}, \texttt{In\_IVFPQ}, \texttt{In\_HNSWFlat}, and \texttt{In\_IVFFlat}.

\begin{figure*}[t]
    \captionsetup{aboveskip=-0.5pt}
    \centering
    \includegraphics[width=\textwidth]{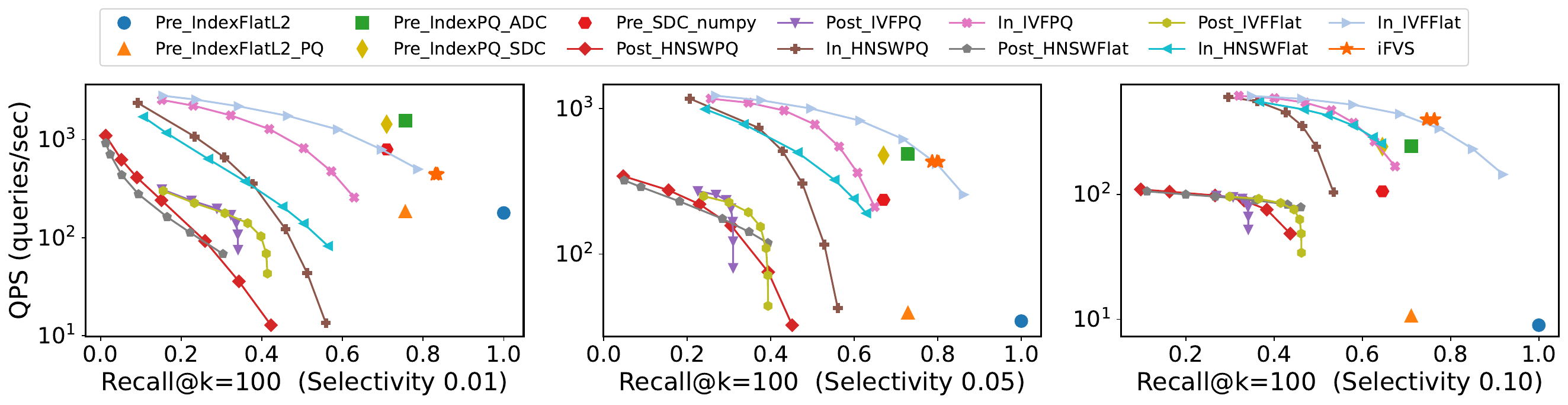}
    \caption{FVS over the SIFT1M dataset.}
    \label{fig:1m-qps-recall}
\end{figure*}
\begin{figure*}[t]
    \captionsetup{aboveskip=-0.5pt}
    \centering
    \includegraphics[width=\textwidth]{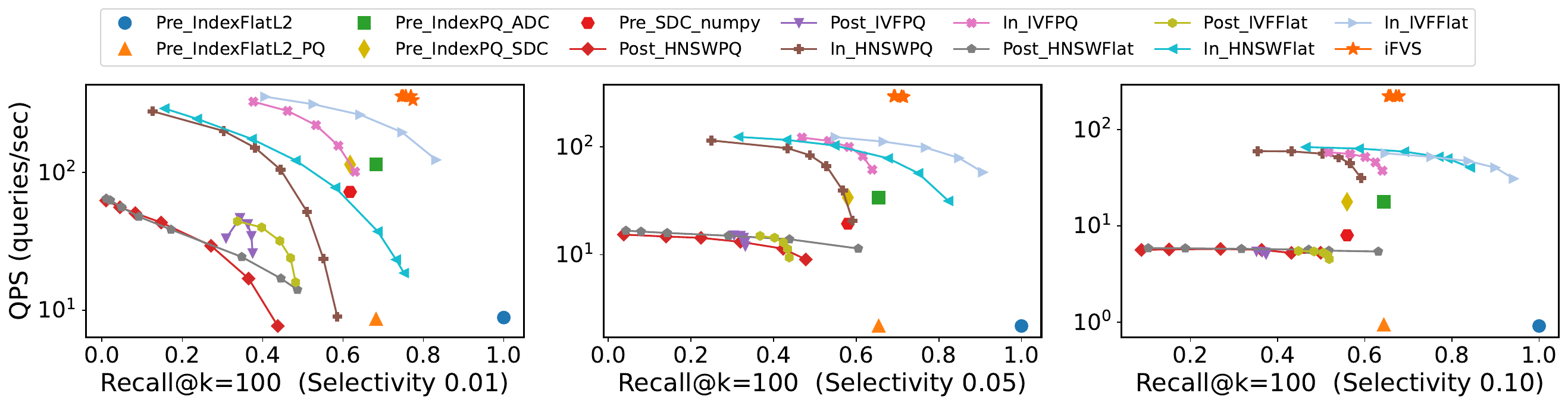}
    \caption{FVS over the SIFT10M dataset.}
    \label{fig:10m-qps-recall}
\end{figure*}
\subsection{Performance Evaluation}
\subsubsection{QPS vs. recall Tradeoff}
~\Cref{fig:1m-qps-recall,fig:10m-qps-recall} show the QPS vs. recall tradeoff on the full SIFT1M, and SIFT10M 
% datasets
% YR: Check if you are okay with it or not
%AAM: Yes, this is better
%\yr{
FVS query workloads, respectively. iFVS splits the FVS query workload into training and test queries. The training queries are used to learn the filter-aware codebook adjustments, while the test queries show the performance of iFVS beyond the queries used for training. The figures report the average recall across both sets, while Table~\ref{tab:recall-membank-bin001} reports them separately.
%}

{\em iFVS} achieves a strong balance between recall and throughput across all selectivity bins. On SIFT1M, it obtains Recall@100 of 0.836, 0.802, and 0.762 at 1\%, 5\%, and 10\% selectivity, respectively, while maintaining 443, 429, and 393 QPS. On SIFT10M, it obtains Recall@100 of 0.774, 0.715, and 0.678, with 334, 291, and 223 QPS. These results show that {\em iFVS} maintains high throughput while improving the ranking quality of compact PQ representations.

\subsubsection{Comparison with PQ-encoded Baselines}
{\em iFVS} consistently improves recall over fixed-codebook PQ baselines. For example, on SIFT10M at 1\% selectivity, Pre\_PQ\_ADC achieves 0.682 Recall@100 and 114 QPS, while {\em iFVS} achieves 0.774 Recall@100 and 334 QPS. Similar trends hold for PQ-SDC and numpy-based SDC baselines. This suggests that adapting the codebook to the FVS query workload improves the accuracy of distance estimation over compressed vectors.

\subsubsection{Comparison with Post- and In-filtering Baselines}
{\em iFVS} dominates all post-filtering baselines across both datasets and all selectivity bins, improving both recall and QPS. This is expected because post-filtering spends search effort on vectors that may not satisfy the predicate. 
{\em iFVS} outperforms all PQ-compressed in-filtering baselines (In\_HNSWPQ, In\_IVFPQ) in both metrics. The exceptions are raw-vector in-filtering methods (In\_IVFFlat, In\_HNSWFlat), which can achieve higher recall at larger selectivities. These methods store full-precision vectors, while {\em iFVS} uses compact PQ codes. Even in these cases, {\em iFVS} remains faster in QPS.

\subsubsection{Effect of Selectivity}
As selectivity increases, the eligible candidate pool becomes larger. This makes ranking more expensive. 
% and more difficult. 
{\em iFVS} degrades gradually under this change. On SIFT1M, recall decreases from 0.836 to 0.762 as selectivity increases from 1\% to 10\%, while QPS decreases from 443 to 393. On SIFT10M, recall decreases from 0.774 to 0.678, while QPS decreases from 334 to 223. Thus, the method remains effective across multiple selectivity bins.

\subsubsection{Effect of Memory-bank Size}
Table~\ref{tab:recall-membank-bin001} studies the effect of memory-bank size $\mathcal{M}$ at 1\% selectivity. Increasing $\mathcal{M}$ improves training recall on both datasets. On SIFT1M, training recall increases from 0.9094 at $\mathcal{M}=4{,}096$ to 0.9709 at $\mathcal{M}=45{,}000$. On SIFT10M, it increases from 0.8195 to 0.8906. However, test recall is highest with the smallest memory bank: 0.7491 on SIFT1M and 0.6706 on SIFT10M. This shows that memory-bank capacity controls a tradeoff between workload specialization and generalization.

\begin{table}[b]
  \centering
  \caption{Recall@100 vs. $\mathcal{M}$ (selectivity = 1\%)}
  \label{tab:recall-membank-bin001}
  \begin{tabular}{r cc cc}
  \toprule
   & \multicolumn{2}{c}{\textbf{SIFT1M}} & \multicolumn{2}{c}{\textbf{SIFT10M}} \\
  \cmidrule(lr){2-3} \cmidrule(lr){4-5}
  Memory Bank & \multirow{2}{*}{Train} & \multirow{2}{*}{Test} & \multirow{2}{*}{Train} & \multirow{2}{*}{Test} \\
  Size, $\mathcal{M}$ & & & & \\
  \midrule
  4{,}096  & 0.9094 & 0.7491 & 0.8195 & 0.6706 \\
  8{,}192  & 0.9270 & 0.7338 & 0.8402 & 0.6582 \\
  16{,}384 & 0.9490 & 0.7212 & 0.8629 & 0.6493 \\
  32{,}768 & 0.9641 & 0.6976 & 0.8868 & 0.6508 \\
  45{,}000 & 0.9709 & 0.7017 & 0.8906 & 0.6572 \\
  \bottomrule
  \end{tabular}
\end{table}

\subsubsection{Index Size and Construction Time}
\begin{figure}[htbp]
    \captionsetup[subfigure]{aboveskip=-0.5pt}
    \captionsetup{aboveskip=-0.5pt}
    \centering
    \begin{subfigure}{0.49\columnwidth}
        \centering
        \includegraphics[width=\linewidth]{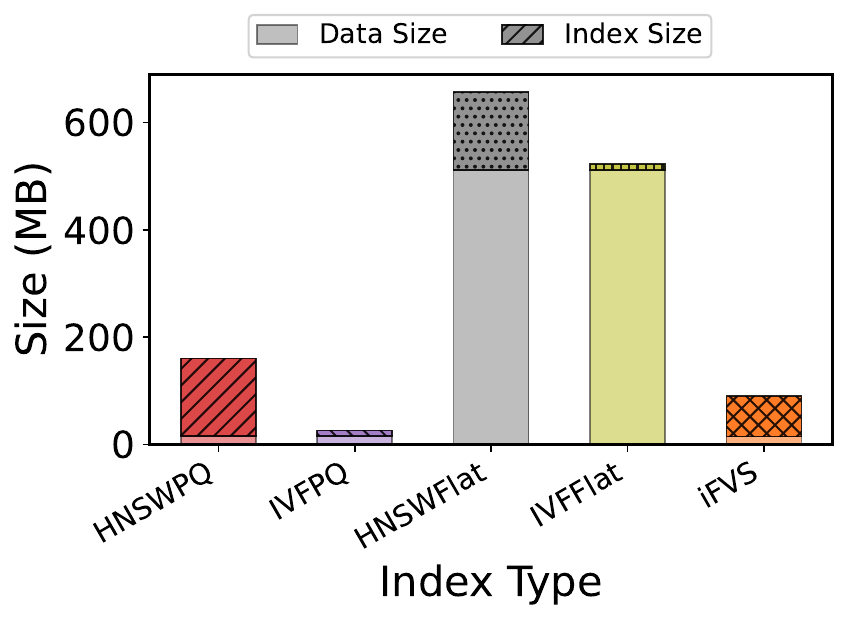}
        \caption{SIFT 1M.}
        \label{fig:1m-idx-size}
    \end{subfigure}
    \begin{subfigure}{0.49\columnwidth}
        \centering
        \includegraphics[width=\linewidth]{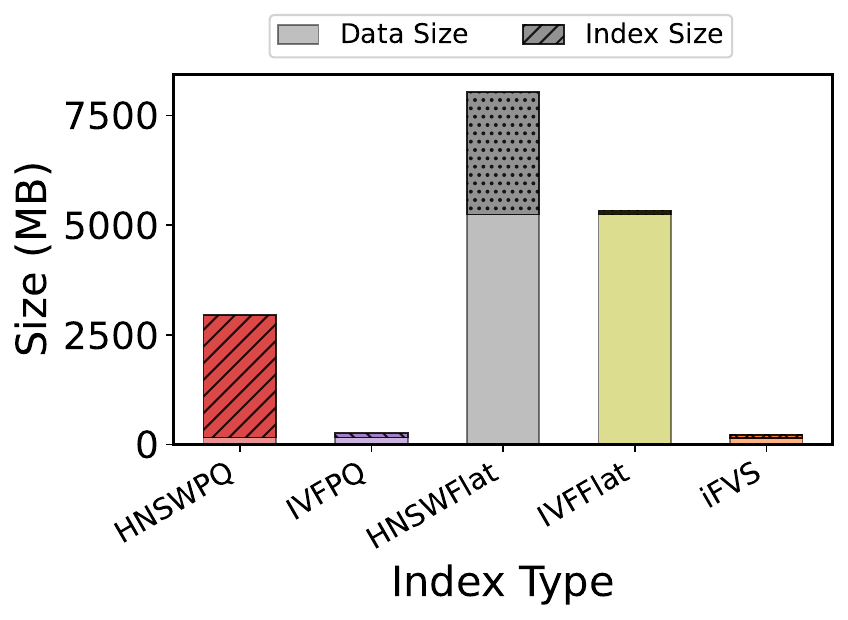}
        \caption{SIFT 10M.}
        \label{fig:10m-idx-size}
    \end{subfigure}
    \caption{Index size comparison.}
    \label{fig:idx-size}
\end{figure}

\begin{figure}[htbp]
    \captionsetup[subfigure]{aboveskip=-0.5pt}
    \captionsetup{aboveskip=-0.5pt}
    \centering
    \begin{subfigure}{0.49\columnwidth}
        \centering
        \includegraphics[width=\linewidth]{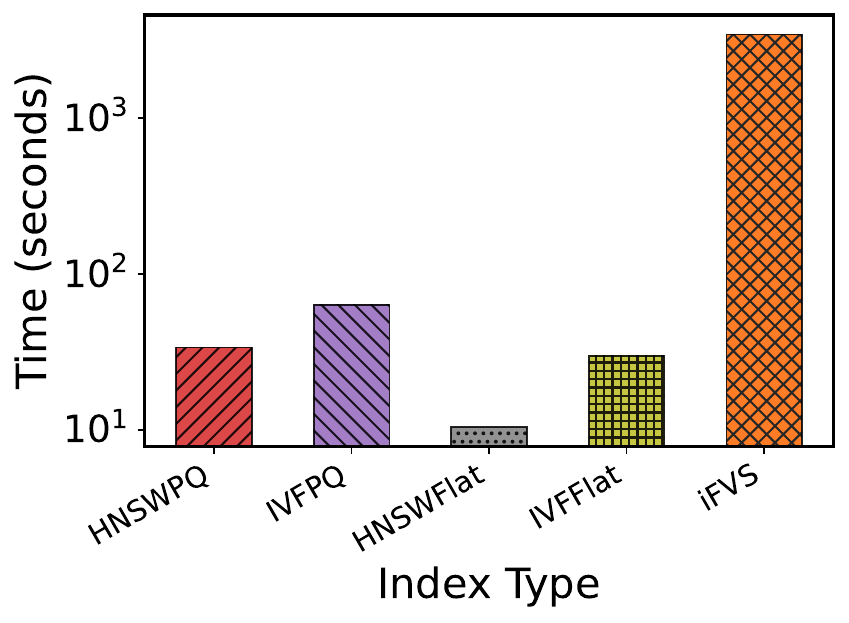}
        \caption{SIFT 1M.}
        \label{fig:1m-con-time}
    \end{subfigure}
    \begin{subfigure}{0.49\columnwidth}
        \centering
        \includegraphics[width=\linewidth]{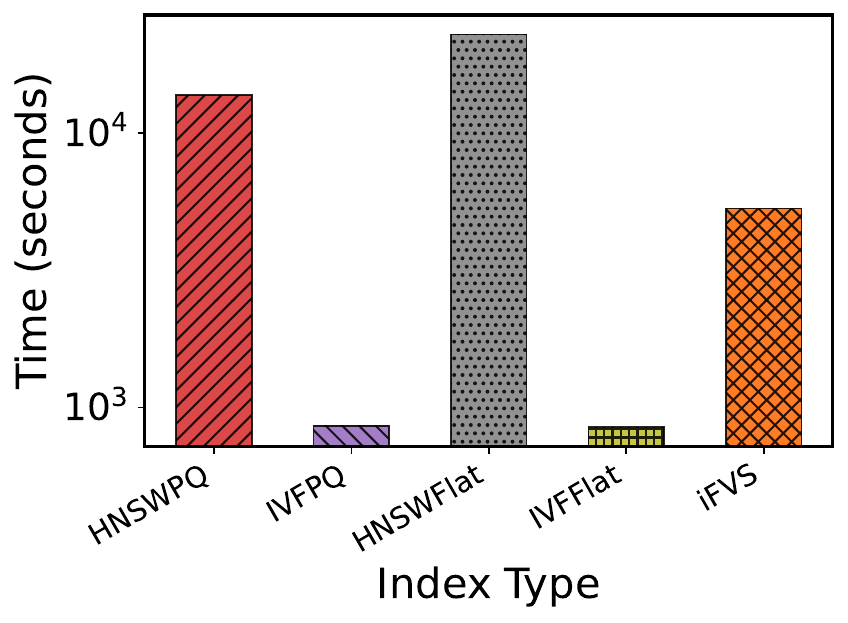}
        \caption{SIFT 10M.}
        \label{fig:10m-con-time}
    \end{subfigure}
    \caption{Index construction time.}
    \label{fig:con-time}
\end{figure}
{\em iFVS} provides compact storage compared with graph-based and raw-vector baselines. On SIFT1M, the total index size is 89.9 MB, which is 1.8$\times$ smaller than HNSWPQ, 7.3$\times$ smaller than HNSWFlat, and 5.8$\times$ smaller than IVFFlat. On SIFT10M, the index size is 220.5 MB, which is 13.4$\times$ smaller than HNSWPQ, 36.4$\times$ smaller than HNSWFlat, and 24.2$\times$ smaller than IVFFlat. The main cost is offline construction time: 57 minutes on SIFT1M and 89 minutes on SIFT10M. This is slower than IVF-based baselines but faster than HNSW-based baselines at 10M scale.

\subsubsection{Scaling from SIFT1M to SIFT10M}
{\em iFVS} scales favorably as the dataset grows by 10$\times$. At 1\% selectivity, QPS decreases from 442.6 on SIFT1M to 333.8 on SIFT10M, a slowdown of only 1.33$\times$. In contrast, Pre\_FlatL2 slows down by 20.2$\times$, and Pre\_PQ\_ADC slows down by 13.6$\times$. 
At 10M scale, {\em iFVS} achieves higher QPS than every evaluated baseline across all selectivity bins. 

Overall, the experiments show that filtered-query-specific codebook generation improves the QPS-recall tradeoff for quantized FVS. {\em iFVS} consistently improves over fixed-codebook PQ baselines, dominates post-filtering baselines, and scales well from 1M to 10M vectors. Its main tradeoffs are offline construction cost and sensitivity to memory-bank capacity.

\section{Conclusion and Future Work}
Instance-optimized FVS is a promising direction for querying large scale vector datasets where quantization is necessary. The proposed {\em iFVS} improves the QPS-recall tradeoff across varying filter selectivities over many of the baselines. Moreover, our experimental evaluation suggests that filtered-query-aware codebook adaptation provides a practical path towards instance-optimized FVS.
In future work, we plan to investigate: (a) the performance of {\em iFVS} on recently proposed filtered query benchmarks~\cite{zhang2026vecbench, lu2026depth, iff2025benchmarking, song2026favor}, (b) new optimization techniques so that the generalization over unseen queries are improved, and (c) self-adjustment of {\em iFVS} in the presence of significant query workload shift.

%\clearpage
\balance
\bibliographystyle{ACM-Reference-Format}
\bibliography{reference}

\end{document}